\documentclass[twocolumn,showpacs,floats,floatfix,superscriptaddress,aps,pra]{revtex4-1}
\usepackage{amsfonts}
\usepackage{amssymb}
\usepackage{amsmath}
\usepackage{graphicx}
\usepackage{bm}
\usepackage{color}
\usepackage{epsfig}
\usepackage{calc}
\usepackage{ifthen}
\usepackage{subfigure}
\usepackage{graphicx}
\usepackage{epstopdf}
\usepackage{epsfig}

\begin{document}

\title{A New Coupling Mechanism Between Two Graphene Electron Waveguides for Ultrafast Switching}
\date{\today }

\begin{abstract}
In this paper, we report a novel coupling between two graphene electron waveguides, in analogy the optical waveguides.
The design is based on the coherent quantum mechanical tunneling of Rabi oscillation between the two graphene electron waveguides.
Based on this coupling mechanism, we propose that it can be used as an ultrafast electronic switching device.
Based on a modified coupled mode theory (CMT), we construct a theoretical model to analyze the device characteristics, and predict that the switching speed is faster than 1 ps and the on-off ratio exceeds $10^{6}$.
Due to the long mean free path of electrons in graphene at room temperature, the proposed design avoids the limitation of low temperature operation required in the traditional design using semiconductor quantum-well structure.
The layout of our design is similar to that of a standard CMOS transistor that should be readily fabricated with current state-of-art nanotechnology.
\end{abstract}

\author{Wei Huang}
\affiliation{SUTD-MIT International Design Centre, Singapore University of Technology and Design, 8 Somapah Road, 487372 Singapore}
\author{Shi-Jun Liang}
\email{shijun\_liang@mymail.sutd.edu.sg}
\affiliation{SUTD-MIT International Design Centre,  Singapore University of Technology and Design, 8 Somapah Road, 487372 Singapore}
\author{Elica Kyoseva}
\email{elkyoseva@gmail.com}
\affiliation{Institute of Solid State Physics, Bulgarian Academy of Sciences, 72 Tsarigradsko Chauss\'ee,
1784 Sofia, Bulgaria}
\author{Lay Kee Ang}
\email{ricky\_ang@sutd.edu.sg}
\affiliation{SUTD-MIT International Design Centre,  Singapore University of Technology and Design, 8 Somapah Road, 487372 Singapore}

\maketitle


\section{Introduction}
Graphene's extremely-high mobility enables the promising application in electronics, but it is notoriously difficult to have a high turn on and off ratio due to the lack of a band gap for single-layer field-effect transistors (FET) \cite{Meric08,Avouris10}.
There have been some extensive research activities in the exploration of novel approaches to achieving high on-off ratios graphene FET, such as utilization of uniaxial strain \cite{Ni08}, lateral confinement \cite{Zhou07} and breaking inversion symmetry \cite{Zhang09}, in order to create a band gap in single-layer graphene.
Another approach is to exploit transport band gap based on few-layer graphene under perpendicular electric field \cite{Xia10}.
A recent approach utilizes the unique electron transport in graphene, called electron optics mechanism in p-n Junctions \cite{Chen16, Sajjad11}.
Unfortunately, the on-off ratio of graphene FET based on above proposals is still lower than $10^6$ even at higher large bias \cite{Britnell2012, Ghobadi2014}.

In this paper, we will explore the coupling between two graphene electronic waveguides based on the coherent quantum mechanical tunneling of Rabi oscillation
The concept of electron waveguide was first proposed in the 90s in analogy to the optical waveguide \cite{Alamo1990, Tsukada1990, Kroemer1989, Alamo1998, Eugster1991,Liang2001, Hrebikova2014}, where the electrons are trapped in a quantum well structure composed of AlGaAs/GaAs material and ballistic electron transport is assumed. Based on this electron waveguide with AlGaAs/GaAs quantum well structure, the researchers described the coupling mechanism between two electron waveguides based on coupled mode theory (CMT) and proposed the concept of electron switching device based on electron waveguide. These types of electron waveguide switches are expected to have ultrafast operating speed.
Recent calculations show that the operating frequency of such a device is up to $0.5\times 10^{12}$ s$^{-1}$ and the maximum coupling energy between the waveguides is 10 meV \cite{Tsukada1990,Kroemer1989}, which corresponds to a very short coupling length of about 280 nm.
Despite these encouraging predictions, it has never been realized experimentally at room temperature probably due to the unjustified assumption of long electron mean free path at the interface of AlGaAs/GaAs material at room temperature that the electrons will suffer inelastic scattering and the coherent phase of the wave packet can not be maintained.
Thus, an electron waveguide based on AlGaAs/GaAs materials at room temperature may not be practical for ultrafast switching application as ultra-low temperature (typically below 4K) is required \cite{Alamo1998}.

This limitation may be lifted with the advances in using graphene electron waveguide \cite{Zhang2009, Hartmann2010, Rickhaus2013, Williams2011, Allen2016}, which have successfully demonstrated ballistic electron transport in a quantum well created on graphene on the length scale of a few $\mu$m.
Tunable Fermi level in the graphene via gating voltage will also offer additional control over the performance of electronic switching device as the tunable barrier height formed at the interface of graphene and different semiconductors (e.g. Si and GaAs) has been confirmed both experimentally and theoretically \cite{SBH,SBH1,Liang2015,Ang2016}, and thus provides another advantage over the fixed band offset at the interface of AlGaAs/GaAs material.
All these new findings may enable a high on-off ratio electronic switching device by using graphene-based electron waveguide.

In this paper, firstly, we introduce coupled mode theory (CMT) to determine the coupling between two parallel graphene electron waveguides, which may serve as a novel ultrafast electronic switching device, in analogy to the optical dual-channel waveguide device. The schematic diagram of the proposed design is shown in Fig.1.
As compared to the AlGaAs/GaAs quantum well structure, there are three main advantages as summarized below.
Firstly, graphene electron waveguide has long electron mean free path, about 4 to 10 $\mu$m \cite{MFPT1} at room temperature, which may be larger than the characteristic length of the device.
Secondly, graphene material has tunable Fermi energy level via gating voltage.
Lastly graphene electron waveguide can operate at room temperature.
According to our calculated results (see Figs. 2 and 3), we predict that the switching time is faster than 1 ps (as shown in Fig.4) and a very high on-off ratio (exceeds $10^{6}$, as shown in Fig.5).

\begin{figure}[hbtp]
\centering
\includegraphics[width=0.5\textwidth]{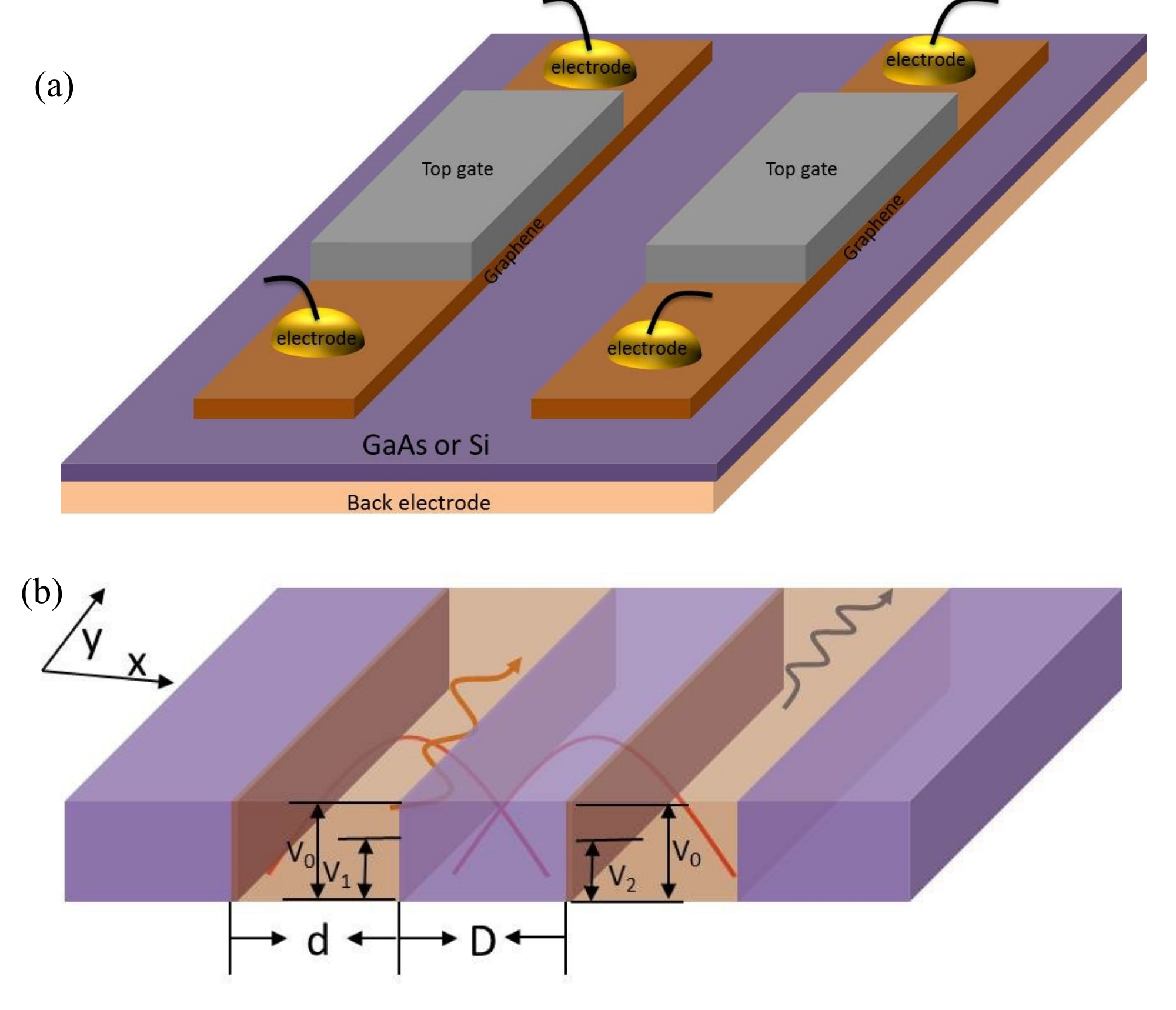}
\caption{(a) The scheme of ultra-fast electron switch based on dual graphene electron waveguide. (b) The corresponding energy band structure of (a). $V_0$ is the Schottky barrier height between graphene and GaAs or silicon material and $V_1$, $V_2$ are the bias voltages applied to modulate the coupling length between the two graphene waveguides. $d$ is the width of the graphene quantum wells and $D$ is the distance between the two graphene electron waveguides.}
\end{figure}


\section{Coupled mode theory of graphene electron waveguide}

Coupled-mode theory (CMT) was initially developed for guided-wave optics to describe the coupling between adjacent optical waveguides, due to the overlap of their evanescent electromagnetic fields.
This allows light to be transferred robustly between the optical waveguides \cite{Yariv1973, Longhi2005, Huang2014}.
By drawing the analogy between the wave nature of electrons (as massless particles) traveling inside the graphene to electromagnetic waves in optical waveguides, CMT is revised to describe the coupling between two parallel graphene electron waveguides as shown in Fig.1.
When the graphene electron waveguides are closely positioned, electrons can be efficiently coupled.
For the proposed electron switching device, we assume that the two bias gate voltages $V_1$ and $V_2$ are equal, thus realizing an electron tunneling version of the Rabi oscillations.
Based on CMT, the coupling length $L$ will depend on a coupling parameter defined as $\Omega=C_{1/2}$ (see definition below) between the two graphene waveguides, namely $L=(2n+1)\pi/ \Omega$, where $n$ is integer, $f_{T}=v_{f}/L$ is the transition frequency, and $v_{f} = 10^6$ m/s is the Fermi velocity of electrons in graphene.

In our two-dimensional (2D) model (see Fig. 1), the proposed ultrafast quantum field-effect transistor (FET) behaves like an electron switching device consisting of two parallel graphene electron waveguides.
The width of the waveguide is $d$ and the separation of the two waveguides is $D$.
Aligned with the standard complementary metal-oxide-semiconductor (CMOS) FET terminology, the two parallel graphene waveguides  can be regarded as a source waveguide and a drain waveguide.
For each graphene waveguide, the two ends are referred as its input and output for the electrical signal.
The ohmic contact between the metal electrodes and the graphene is assumed to enable sufficient electron injection into the waveguides.

When a small voltage is applied to the left graphene waveguide, an electrical current can be measured at the outputs of both graphene waveguides.
The gate voltages ($V_1$ and $V_2$) on each graphene waveguide are used to tune the Fermi level of each channel independently.
Consequently, the Schottky barrier height ($V_0$) and thickness layer ($D$) between the graphenes can be modulated.
If the Schottky barrier height and the effective gap spacing between the two graphene waveguides are sufficiently small, the evanescent wave of the injected electrons in the source waveguide can tunnel into the drain waveguide with some probability as represented by the red curve in Fig. 1b, which can be defined as an on-state.
When the Schottky barrier height and effective gap spacing become large due to the applied gate voltage, the tunneling probability of electron tunneling from source waveguide to drain waveguide is extremely small, which can be defined as an off-state (In our example, we set off-state of gate voltages as $V_1 = V_2 =$ 300 meV). Note that the phase of the electrons is maintained during the tunneling process.
Quantum mechanically, the injected electrons in the source waveguide can be detected at the drain waveguide with a probability equal to 1 [see black curved arrow in Fig. 1 (b)] at a certain transfer length $L$ .

In our model, we assume that a quantum well is created at the interface between the graphene and GaAs.
The electrons in the graphene waveguides are confined along with the $x$-direction, and they unbounded in the $y$-direction.
The total electron wave function is the superposition of all possible quantum eigenstates.
We make the notations $\Psi_1(x,y)$($\Psi_2(x,y)$) as the electron wave function of source (drain) graphene waveguide, which is written as
\begin{equation}
\begin{aligned}
\Psi_1(x,y)=\sum_m a_{1m}(y) u_{1m}(x) \exp(-i \beta_{1m} y), \\
\Psi_2(x,y)=\sum_n a_{2n}(y) u_{2n}(x) \exp(-i \beta_{2n} y),
\end{aligned}
\end{equation}
where $a_{1m}$, $a_{2n}$ are $m^{th}$, $n^{th}$ modes with respect to source/drain graphene waveguide.
Here, $u_{1m}(x)$, $u_{2n}(x)$ are the mode profiles of the wavefunctions determined by electron eigenstates in the quantum well, where $m,n$ are the mode indexes. $\beta_{1m} = k_1 \sin(\theta_m)$ and $\beta_{2n} = k_2 \sin(\theta_n)$ are the respective propagation constants of the $m^{th}$ and $n^{th}$ modes in graphene electron waveguide with respect to source/drain graphene waveguide.
The parameters $\theta_{m}$ and $\theta_{n}$ describe the injection electron angles with respect to the corresponding $m^{th}$ and $n^{th}$ modes in the source/drain graphene waveguide, $k_1$ and $k_2$ are the wave vectors of electron injected in the source/drain graphene waveguide: $k_1=(E-V_1)/ \hbar v_F$ and $k_2=(E-V_2)/ \hbar v_F$. The wave vector in the barrier material is $k_0=\sqrt{-2m_{\text{eff}}(E-V_0)/\hbar^2}$, where $E$ is the electron energy and $m_{\text{eff}}$ is the electron effective mass in the semiconductor material.

According to Myoung's paper \cite{Myoung2011}, the electron's wave function behaves as a plane wave along with the $y$ direction and Eq. (1) can be rewritten as
\begin{equation}
\begin{aligned}
\Psi_1(x,y)=\sum_m a_{1m}(y) \psi_{1m},  \\
\Psi_2(x,y)=\sum_n a_{2n}(y)  \psi_{2n},
\end{aligned}
\end{equation}
where $\psi_{1m}$=$u_{1m}(x) \times \exp(-i \beta_{1m} y)$ and $\psi_{2n}$=$u_{2n}(x) \times \exp(-i \beta_{2n} y)$.
In the $y$ direction, the motion of the electrons in the graphene electron waveguides is described by the 1D free electron Dirac equation, which allows us to decouple the wavefunctions into two sublattices: A and B wave functions (see the Appendix for details).
Finally, two Helmholtz-like equations in the $y$ direction are obtained as
\begin{equation}
\begin{aligned}
\dfrac{\partial^2}{\partial y^2} \psi_{1m} + \beta_{1m}^2 \psi_{1m}=0, \\
\dfrac{\partial^2}{\partial y^2} \psi_{2n} + \beta_{2n}^2 \psi_{2n}=0.
\end{aligned}
\end{equation}

Based on the CMT model \cite{Yariv1973} we can manipulate the Helmholtz equations to obtain
\begin{equation}
\begin{aligned}
\dfrac{\partial^2}{\partial y^2} \Psi_{1m}(x,y) + \beta_{1m}^2 \Psi_{1m}(x,y) = -(k_{2}^2 - k_0^2)\Psi_{2n}(x,y), \\
\dfrac{\partial^2}{\partial y^2} \Psi_{2n}(x,y) + \beta_{2n}^2 \Psi_{2n}(x,y) = -(k_{1}^2 - k_0^2)\Psi_{1m}(x,y),
\end{aligned}
\end{equation}
which are consistent with previous optical waveguide coupled equations \cite{Saleh1991} and the electron waveguide coupled equations \cite{Alamo1990}.

By substituting Eq.(1) into the Eq.(4), and considering $\psi_{1m}$ and $\psi_{2n}$ obeying Eq.(3), we apply the slowly envelope varying approximation \cite{Saleh1991} to Eq. (4), such that $\dfrac{d^2 a_1}{dy^2} \ll \dfrac{d a_1}{dy}$ and $\dfrac{d^2 a_2}{dy^2} \ll \dfrac{d a_2}{dy}$, and obtain the following coupling equations:
\begin{equation}
\begin{aligned}
\dfrac{d a_{1m}}{dy} e^{-i \beta_{1m} y} = -i C_{12} a_{2n} e^{-i \beta_{2n} y}, \\
\dfrac{d a_{2n}}{dy} e^{-i \beta_{2n} y} = -i C_{21} a_{1m} e^{-i \beta_{1m} y}.
\end{aligned}
\end{equation}
Here, $C_{12}$ and $C_{21}$ are the coupling coefficients, which are given by,
\begin{equation}
\begin{aligned}
C_{12} = \dfrac{1}{2} \dfrac{k_{2}^2 - k_0^2}{\beta_{1m}}  \int^{+\infty}_{-\infty} u_{1m}(x) u_{2n}(x) dx, \\
C_{21} = \dfrac{1}{2} \dfrac{k_{1}^2 - k_0^2}{\beta_{2n}}  \int^{+\infty}_{-\infty} u_{1m}(x) u_{2n}(x) dx.
\end{aligned}
\end{equation}
Finally, Eq.(5) can be rewritten as a Schr\"odinger-like equation of a two-level system,
\begin{equation}
i\dfrac{d}{d y}
\begin{bmatrix}
a_{1m} \\
a_{2n}
\end{bmatrix}
= \begin{bmatrix}
0 & C_{12} e^{i y \Delta} \\
C_{21} e^{-i y \Delta} & 0
\end{bmatrix} \begin{bmatrix}
a_{1m} \\
a_{2n}
\end{bmatrix},
\end{equation}
where $\Delta = \beta_{1m} - \beta_{2n}$.
It is important to note that Eqs. (6) and (7) are the master coupling equation for determining the coupling between $m^{th}$ mode and $n^{th}$ mode in source/drain graphene electron waveguide studied in this paper.
They are valid for the coupling of any dual-graphene electron waveguide device.

Before presenting the results, it is important to note that graphene is used as the electron waveguide channel and GaAs is the material for the potential barrier.
The electrons inside the graphene waveguide channels are described by Dirac equation and Schr\"{o}dinger equation is used to describe electrons dynamics in the potential barrier (GaAs).
Previous work on single graphene electron waveguide channel \cite{ Zhang2009, Hartmann2010, Yuan2011, Myoung2011} has shown that the evanescent electron wave will exponentially decay outside of graphene electron waveguide.
According to Hartmann's \cite{Hartmann2010} and Xu's \cite{Xu2015} papers, there are two kinds of stable modes in graphene electron waveguide dependent on two sub-lattice modes: $\psi_A$ and $\psi_B$, where $\psi_A+i\psi_B$ is anti-symmetric and $\psi_A-i\psi_B$ is symmetric.
The coupling between the same anti-symmetric functions, and the coupling between the antisymmetric and symmetric functions is, respectively, 0.24 meV and 0.14 meV, which are much smaller than the coupling between the two symmetric functions (1.58 meV).

\begin{figure}[hbtp]
\centering
\includegraphics[width=0.5\textwidth]{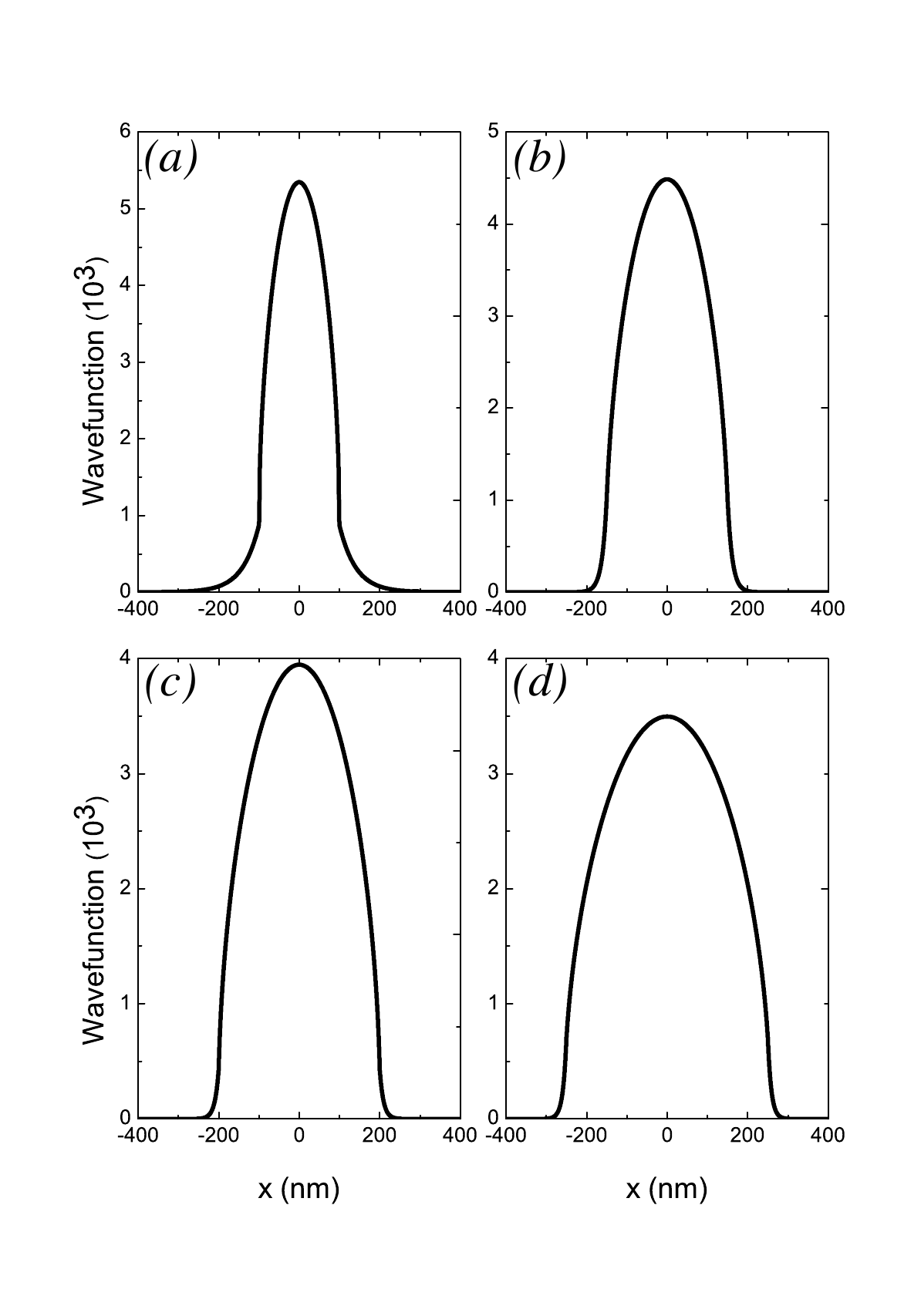}
\caption{The symmetrical wave modes of first mode $u_{1}=u_{2}=\psi_A-i \psi_B$ in the graphene electron waveguide for different widths $d$ of the quantum well. Schottky barrier height between graphene and AlGaAs/GaAs is 500 meV and $V_1=V_2$=450 meV, so that the effective barrier height between the two graphene waveguides is 50 meV and $k_1 d=4.96 \pi$. (a) d=200 nm, $k_{1x} d = 2.9465$, $\theta=79.1^\circ$, (b) d=300 nm, $k_{1x} d = 3.0243$, $\theta=78.809^\circ$,  (c) d=400 nm, $k_{1x} d = 3.1165$, $\theta=78.463^\circ$, (d) d=500 nm, $k_{1x} d = 3.0554$, $\theta=78.669^\circ$.}
\end{figure}

Thus, we only consider the coupling of two symmetric functions in this paper.
In Fig. 2, we show the symmetric wave function of the graphene electron waveguide at the different width of the waveguide: $d$ = 200, 300, 400, and 500 nm.
The probability current density through the interface of graphene and GaAs is conserved \cite{Dragoman2013}.
Unless specified, the default parameters are $V_0$=500 meV, $V_{1}= V_{2}$=450 meV, and $k_1 d$= 4.96$\pi$ $k_{1x}=k_1 \cos(\theta_m)$ is defined as the wave vector in the $x$ direction.

The coupling strength between two parallel graphene electron waveguides is shown in Fig. 3, where we have assigned $k_1 d = 4.96 \pi$ and study the dependence as a function of spacing ($D$) between the waveguides and width of the waveguide $d$.
From Fig. 3, the coupling strength is decreasing with increasing spacing of $D$ from about 0 to 120 nm.
The coupling also increases with decreasing $d$ from 500 nm to 250 nm.
For smaller $d$ = 250 nm or smaller, the mode profile becomes flat, and the overlapping calculated by $\int^{+\infty}_{-\infty} u_{1m}(x) u_{2n}(x) dx$ initially decreases until reaching a critical point, which is determined by the energy and incident angle as well as barrier height of injected electron. Finally we show a contour plot of the transition frequency $f_{T}$ as a function of the spacing $D$ and $d$ in Fig. 4.

\begin{figure}[hbtp]
\centering
\includegraphics[width=0.5\textwidth]{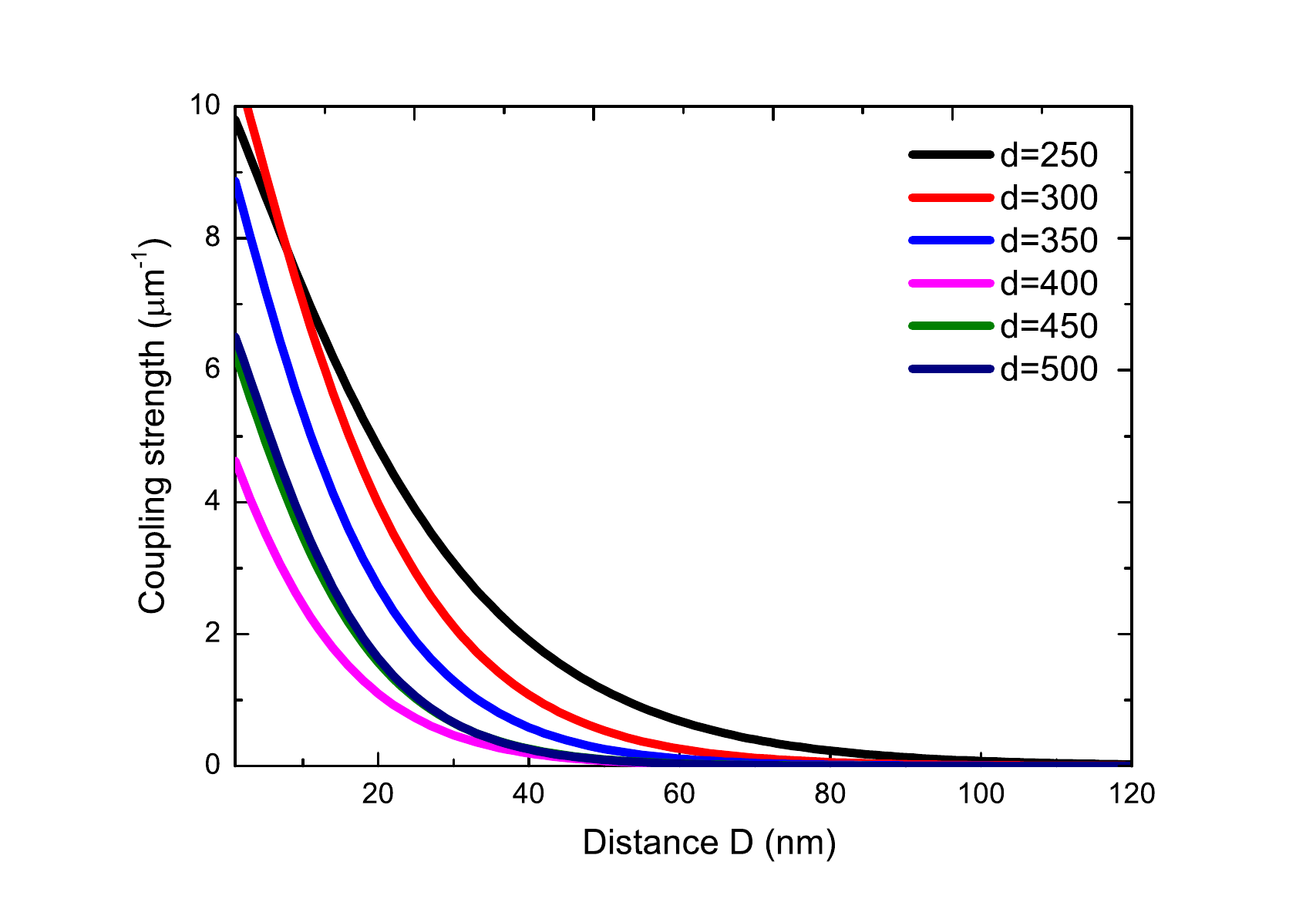}
\caption{The coupling strength between two parallel graphene electron waveguides with fixed $k_1 d = 4.96 \pi$ as a function of spacing $D$ between graphene electron waveguides at the different width of the waveguide $d$.}
\end{figure}

Compared with prior electron switching devices based on AlGaAs/GaAs materials, our proposed design based on graphene waveguides will have larger coupling energy (around 15 meV) and faster operating frequency ($ <$ 1 ps in terms of time scale)
as shown in Fig. 4.
Most importantly, our proposed device is able to operate at room temperature due to the long mean free path of electrons in the graphene as compared to AlGaAs/GaAs material.

\begin{figure}[hbtp]
\centering
\includegraphics[width=0.5\textwidth]{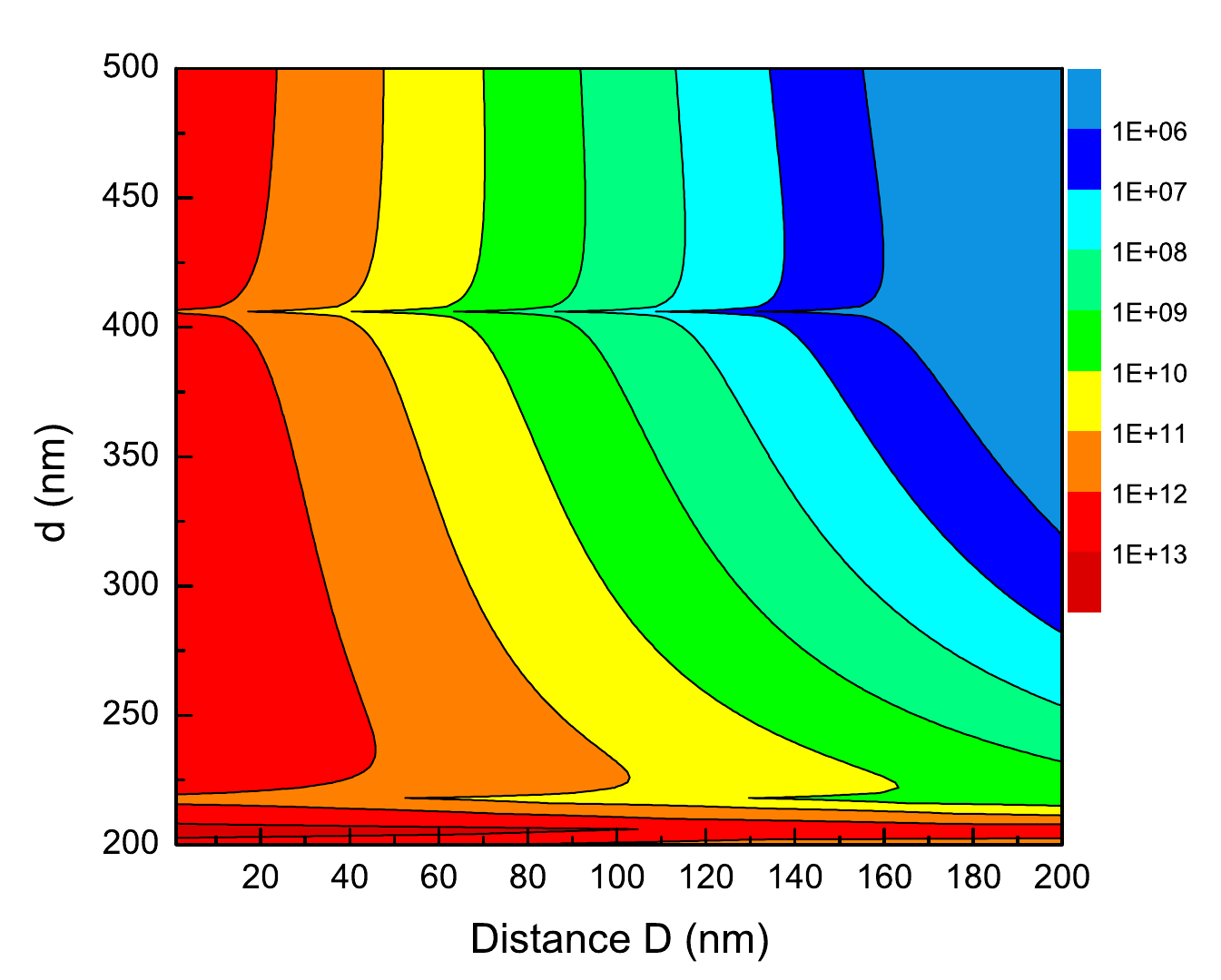}
\caption{Contour plot of complete transfer frequency as a function of the width $d$ of quantum well and distance $D$ between two graphene electron waveguides. We set $k_1d=4.96 \pi$ and Schottky barrier height between graphene and GaAs is 500 meV.  $V_1=V_2=450$ meV. }
\end{figure}

\section{Discussions}

Previous graphene field effect transistors have the low on-off ratio at room temperature \cite{Schwierz2010}.
In spite of many efforts to improve their designs \cite{Britnell2012, Ghobadi2014, Wang2015, Katkov2014}, the on-off ratio of the graphene field effect transistors is still below $10^6$ even at a relatively high voltage of the order of 5 V \cite{Britnell2012, Ghobadi2014}.
From the results obtained above in this paper,  our proposed design can behave as a field effect transistor, in which electrons coupling between the two waveguides can be controlled by the gate voltage.
Typical values of the on-off ratio is presented in Fig. 5, in which we use $D$ = 50 nm and $d$ =200 nm and $k_1d=4.96\pi$.
From the figure, we see that the on-off ratio can exceed $10^6$.

\begin{figure}[hbtp]
\centering
\includegraphics[width=0.5\textwidth]{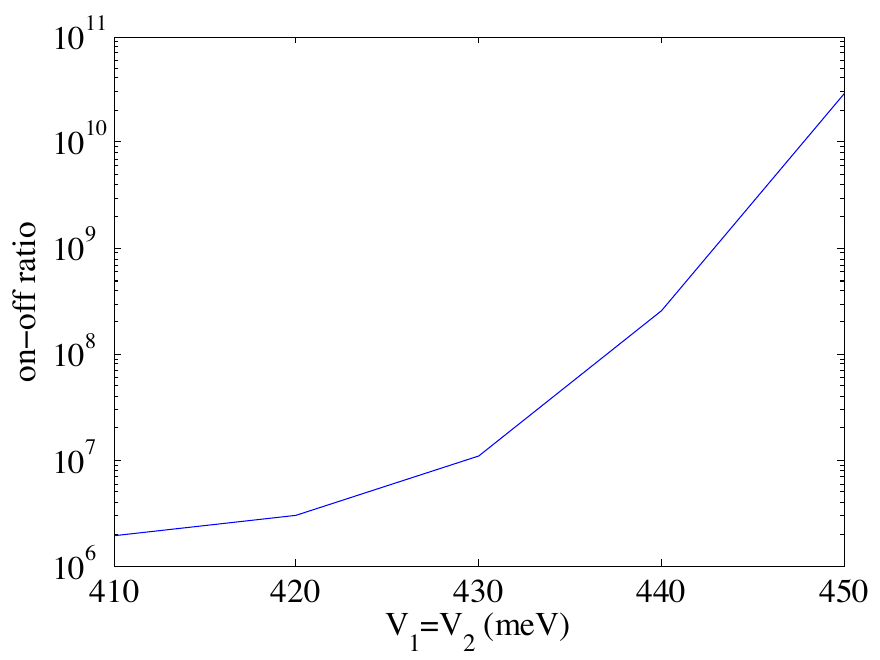}
\caption{ On-off ratio with varying $V_1$=$V_2$ from 410 meV to 450 meV. We set $d=200$ nm and the energy of injection electron is $E=498.28$ meV. Schottky barrier height between graphene and GaAs to be 500 meV. The device length set to be 654nm, which is the coupling length $L$ for $V_1=V_2$=450 meV. }
\end{figure}

From solving Eqs. (6) and (7) numerically, we present the first and second coupling modes of the graphene waveguide in Fig. 6 for $d$ = 200 nm and $D$ = 50 nm
For simplicity, the detuning parameter is set at $\Delta =0$ and thus $\Omega=C_{12}=C_{21}$.
The Schottky barrier height between the graphene and the GaAs is 0.5 eV \cite{SBH1}.
In general, higher barrier height will allow for a lower coupling strength $\Omega$, and leads to a longer coupling length.
We also assume that the symmetrical wave mode $\psi_A-i \psi_B$ is the initially populated mode in the source graphene waveguide, which can be coherently excited by the high-quality contact between the metal electrode and graphene.

\begin{figure}[hbtp]
\centering
\includegraphics[width=0.5\textwidth]{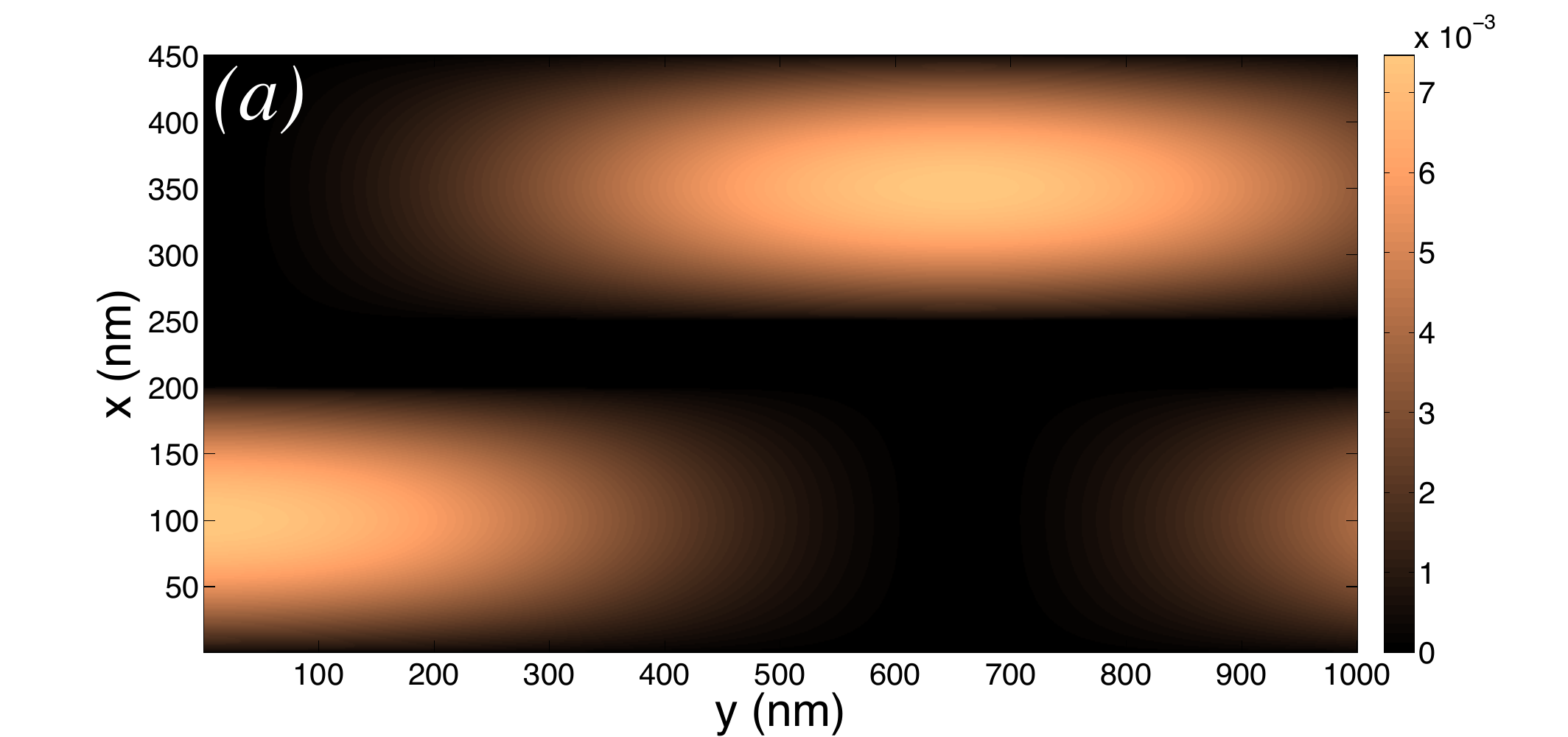}
\includegraphics[width=0.5\textwidth]{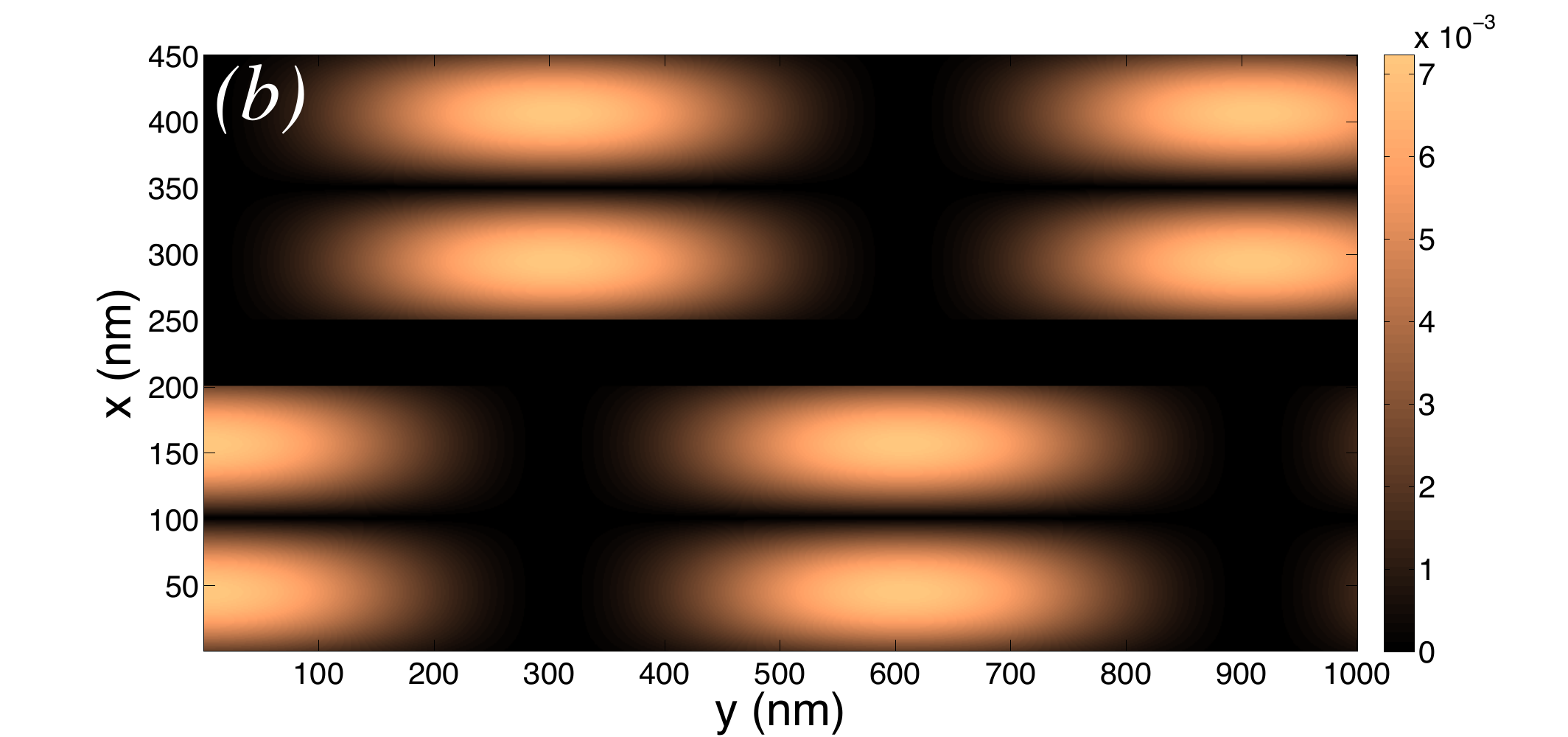}
\caption{Electron wavefunction probability for a system of two coupled graphene electron waveguides. The parameters are set as follows: $d =$  200 nm, $D = $ 50 nm, $k_1d = 4.96 \;\pi$ and the height of the barrier between graphene and GaAs (Si) is 500 meV and $V_1=V_2=$ 450 meV. In the upper frame (a) we consider coupling between the first modes of the graphene waveguides, while in the lower frame (b) the coupling is between the corresponding second modes. }
\label{Fig4}
\end{figure}

Figure 6 clearly shows a demonstration of Rabi oscillations of the electrons probability amplitudes between the two graphene waveguides, in analogy to the dynamics of a two-level quantum mechanical system.
The population of the different modes of the graphene waveguides depends on the gate-controlled guiding of the electrons into the source waveguide.
If the electron injected into the source graphene waveguide has a wave packet perfectly matched at a certain waveguide eigenstate, only that specific mode will be excited.
The figure also demonstrates the electron coupling transfer length $L$ up to 1000 nm is possible.
As an example, the coupling between the first modes in quantum optical waveguides, the coupling length is $L= 654$ nm with a transfer frequency of $f_T = 1.53\times 10^{12}$ s$^{-1}$ (or 0.65 ps), which is very fast (less than 1 ps).

These results show that the proposed electron switching device based on graphene electron waveguides has considerable advantages compared to similar devices using the conventional semiconductors materials in terms of operating speed, compactness and operating at room temperature.
Note we do not consider higher-order coupling between the first mode in source waveguide and the second mode in the drain waveguide.
The reasons are two folds: firstly, the coupling between the first mode (in the source waveguide) and second modes (in the drain waveguide) is asymmetric leading to very weak coupling strength (0.12 meV).
Secondly, the preparation of modes in the waveguides (source or drain) depends on specified electron injection angles, so it is difficult to implement experimentally the coupling between these two modes.
Interestingly, the coupling transfer length will be reduced to $L$ = 302 nm and the transfer frequency will increase to $f_T = 3.3\times 10^{12}$ s$^{-1}$ (0.3 ps) for coupling only between the second modes in the graphene electron waveguide, as shown in the Fig.6 (b).

CMT is a very powerful method to compute coupling strength, for which the only assumption is that two mode profiles would not have changed during overlapping of two evanescent electric fields of two optical waveguides \cite{Yariv1973,Haus91}.
The CMT not only works well for the electric field, also for electron wave function described by quantum mechanics. For example, the coupling mechanism of superconducting Josephson junctions can be described by CMT, as the evanescent wave function of electrons in two superconductors has overlapping \cite{Raghavan99}; Similarly, the CMT has also been utilized to describe the coupling between electronic waveguides based on AlGaAs/GaAs materials. Furthermore, the previous work on graphene electronic waveguide has already shown that wave function of electrons has the similar mode as the optical waveguide \cite{Hrebikova2014, Zhang2009, Hartmann2010, Rickhaus2013, Williams2011}, [see electron wave function modes in graphene (Fig. 2)]. Based on these arguments above, we believe that the CMT is valid for the description of the coupling mechanism between two parallel graphene electron waveguides.

There have been various papers \cite{Gorbachev2012, Schutt2013,Tse07,Kim2011} pointing out that Coulomb drag effect is very profound near the Dirac cone, which is induced by strong electron-electron interaction between two horizontal parallel graphene sheets separated by dielectric materials.
However the graphene layout in our proposed device is coplanar parallel instead of horizontal parallel.
Besides we focus on the condition that the Fermi level is far from the Dirac point.
From the following simple analysis of Coulomb drag effect, we find that the Coulomb drag effect plays a minor role in our model.

When a current $I_{active}$ flows on the active layer of graphene, it will induce a voltage $V_{passive}$ on the passive layer of graphene, due to Coulomb drag effect, which can be measured by drag resistivity $\rho_{drag} = V_{passive} / I_{active}$.
To numerically estimate the effect of the drag effect in our system, the drag resistivity $\rho_{drag}$ based on Boltzmann transport formalism and the random phase approximation is used \cite{Tse07,Kim2011}, which is

\begin{equation}
\rho_{drag} = - \frac{1}{16 \pi k_B T} \sum_{q} \int_{0}^{\infty} dw \dfrac{\Gamma_1(q,w) \Gamma_2(1,2)|U_{1,2}(q,w)|^2}{\sinh^2(\hbar w/ 2 k_B T) \sigma_{L_1} \sigma_{L_2}} .
\end{equation}

Here, subscripts 1 and 2 are the labels for the two single-layer graphenes and $\Gamma_{i}$ is the nonlinear susceptibility of layer $i$. $U_{1,2}$ is the screened interlayer potential in the random phase approximation.  $\sigma_{L_i}$ is the longitudinal conductivity of the layer $i$ and $k_B$ is the Boltzmann constant.
The current flow $I_2$ in the passive graphene layer caused by coulomb drag can be estimated as, $I_{passive} = \rho_{drag} I_1 \sigma d$, where $\sigma=\frac{4 e^2}{h} \frac{E_F \tau}{2 \hbar}$ denotes the conductivity of the passive graphene layers \cite{adam2007}, $d$ is the width of our graphene layer and $\tau$ is the scattering time.
For a typical set of parameters: $D=50$ nm, $T=300$ K, $d=200$ nm and $E_F = 450$ meV, we get the ratio of $I_{passive}/I_{active}$ is 0.18 \%, which is extremely small.
Thus the Coulomb drag effect in our model is negligible and will not greatly affect the calculations reported in this paper.

Due to the electrons scattering, the phase of electrons will be randomly changed. However, we can estimate the phase coherence length of electron in semiconductor at room temperature. If our device setting up of distance between two graphene (D = 50 $nm$) is smaller than the phase coherence length of electron, we can ignore the phase changing of the electrons.  The phase coherence length of electron can be described by the equation, $L_{\phi}= \sqrt{2}(\frac{D_{coe}^2 m_{\text{eff}} L}{\pi k_B T})$ \cite{Altshuler1982, Aleiner1999}, where $D_{\text{coe}}$ is the diffusion coefficient and $k_B$ is the boltzmann constant. At the ultra-low temperature, the phase coherence length can be up to 18 $\mu m$ \cite{Topinka2001}. By substituting room temperature (T = 300 $K$) and geometry setting up of our device (L = 654 $nm$) into the equation, we can obtain the phase coherence length is around 80 $nm$, where the diffusion coefficient is given by Einstein relation $D_{\text{coe}} = \mu k_B T / e$. Therefore, our estimated phase coherence length $L_{\phi} $ =80 $nm$ is larger than distance between two graphene (D = 50 $nm$) and we can ignore the phase changing of the electron in our configuration. Furthermore, this result is corresponding to electron-phonon collisions rate is around $10^{10}$, where our switch time is 1 $ps$ and this electron-phonon collisions rate is consisting with experiment date \cite{Muhammad2017}.

Conventional graphene electrical switches are mainly based on the field effect transistor \cite{Meric08, Avouris10, Ni08, Zhou07}, which utilizes the tunneling effect in the conduction channel between the source and drain electrodes.
Unlike those conventional switches, the proposed electrical switch is not based on this traditional principle although its configuration is similar to conventional field effect transistors.
The operating principle of our electrical switch depends on the electron wave coupling between the source graphene electron waveguide and the drain waveguide, which exhibits Rabi oscillations.

To check the sensitivity of the device against the fluctuations in the chemical potential, we assumed that the variation of the chemical potential is about $1$ meV.
Our calculations show that the resulting variation in the on-off ratio is less than $10\%$.
Large chemical potential can ensure the stable operation of the proposed electrical switch without being affected by external environmental factors.
The variation of $1$ meV in the Fermi level may be justified by the large Fermi level.
Previous paper \cite{Hajaj2013} has reported that fluctuations in the carrier density induced by impurities in the environment or other factors disappear and chemical potential distribution becomes homogeneous when Fermi level is far from the neutral point in the graphene.
If it is near the Dirac cone of the graphene, while fluctuations in the chemical potential may affect the operation of our device.

Additionally, the distance between two electron waveguide sets to be larger than 50 nm, so that the coupling between the two graphene sheets is too weak to open a gap in the graphene electronic bandstructure. Both theory and experiments have an already confirmed that a very weak coupling between two separated graphene sheets cannot lead to a band gap if the gap distance between two graphene sheets is beyond 10 nm \cite{Magda2014,Zhao2011}. Importantly, it is worth to emphasize that the proposed device operates at room temperature \cite{adam2007,qiuzhi2011}, as a result, the exchange interaction between the electrons will be smeared out by large temperature fluctuations.

\section{Conclusion}
In summary, we have studied the coupling between two graphene electron waveguides by utilizing coupled mode theory (CMT).
Based on this coupling mechanism, we have proposed an ultrafast electron switching device based on a dual-graphene-electron-waveguides structure.
The performance of this design is analyzed by using a modified couple model theory together with solving the Dirac and Schr\"{o}digner equations.
Based on our model, it is possible to achieve higher operating frequency (less than 1 ps time scale) with short coupling transfer length at room temperature.
This performance is better than the conventional design in using traditional AlGaAs/GaAs quantum well structure, which will also require very low operating temperature in order to have a long electron mean free path to ensure ballistic electron transport.
The proposed design can be realized using current state-of-art-technology, for example single graphene electron waveguide channels have recently been fabricated with the gate-controlled guiding of electrons \cite{Williams2011}.
Our design resembles the traditional CMOS transistor in its layout, and thus its fabrication is feasible and it has the potential to contribute to the rapid development of quantum circuits and other integrated electron devices.

\section{Acknowledgment}
This work is partially supported by the Singapore ASTAR AME IRG (A1783c0011) and USA Air Force Office of Scientific Research (AFOSR) through the Asian Office of Aerospace Research and Development (AOARD) under Grant No. FA2386-17-1-4020.
EK acknowledges financial support from the European Union's Horizon 2020 research and innovation programme under the Marie Sk\l odowska-Curie grant agreement No 705256 --- COPQE.

\section{Appendix}
Firstly, we show that the Helmholtz-like equations (Eq. (3)) can be described as source/drain graphene electron waveguide in the $y$ direction propagation.
We take the source graphene electron waveguide as an example in our derivation and only consider single mode.
Two sublattices A and B's wavefunction of source graphene electron waveguide refer to as $\psi_A$ and $\psi_B$.
Electron is unbounded in the $y$-direction and electron's energy is $E\sin(\theta_1)$, where $\theta_1$ is electron injection angle of source graphene electron waveguide.
Based on the Dirac equations for free particles, we have

$$
\begin{aligned}
-i\hbar v_F \dfrac{\partial}{\partial y} \psi_B  = (E-V_1) \sin(\theta_1) \psi_A, \\
-i\hbar v_F \dfrac{\partial}{\partial y} \psi_A  = (E-V_1) \sin(\theta_1) \psi_B.
\end{aligned}
\eqno{(A1)}
$$

Equation (A1) can be rewritten as $i \dfrac{\partial}{\partial y} \psi_B + \beta_1 \psi_A = 0$ and $i \dfrac{\partial}{\partial y} \psi_A + \beta_1 \psi_B = 0$, with $\beta_1 = (E-V_1) \sin(\theta_1)/ \hbar v_F$.
By taking the derivative on both side of the equations to decouple $\psi_A$ and $\psi_B$, we obtain the Helmholtz-like equation for the source graphene electron waveguide: $\dfrac{\partial^2}{\partial y^2} \psi_A + \beta_1^2 \psi_A =0$ and  $\dfrac{\partial^2}{\partial y^2} \psi_B + \beta_1^2 \psi_B =0$.

For source graphene electron waveguide, we only consider $\psi_1=\psi_A - i\psi_B$, and the governing equation becomes $\dfrac{\partial^2}{\partial y^2} \psi_1 + \beta_1^2 \psi_1 =0$.
Repeat the same procedure, the Helmholtz-like equation for the drain graphene electron waveguide is obtained as $\dfrac{\partial^2}{\partial y^2} \psi_2 + \beta_2^2 \psi_2 =0$.

Below we show the derivation of Eq. (5) from Eq. (4) in the main text.
\begin{widetext}
$$
\begin{aligned}
\dfrac{\partial^2 \Psi_1}{\partial y^2} + \beta_1^2 \Psi_1 = \dfrac{\partial^2 a_1}{\partial y^2} u_{1}(x) \exp(-i \beta_{1} y) - 2i \beta_1 \dfrac{\partial a_1}{\partial y} u_{1}(x) \exp(-i \beta_{1} y) - \beta_1^2 a_1 u_{1}(x) \exp(-i \beta_{1} y) + \beta_1^2 a_1 u_{1}(x) \exp(-i \beta_{1} y),\\
\dfrac{\partial^2 \Psi_2}{\partial y^2} + \beta_2^2 \Psi_2 = \dfrac{\partial^2 a_2}{\partial y^2} u_{2}(x) \exp(-i \beta_{2} y) - 2i \beta_2 \dfrac{\partial a_2}{\partial y} u_{2}(x) \exp(-i \beta_{2} y) - \beta_2^2 a_2 u_{2}(x) \exp(-i \beta_{2} y) + \beta_2^2 a_1 u_{1}(x) \exp(-i \beta_{1} y).
\end{aligned}
\eqno{(A2)}
$$
\end{widetext}

In Eq. (A2), we apply slowly envelope varying approximation \cite{Saleh1991}, say $\dfrac{d^2 a_1}{dy^2} \ll \dfrac{d a_1}{dy}$ and $\dfrac{d^2 a_2}{dy^2} \ll \dfrac{d a_2}{dy}$. Slowly envelope varying approximation (SEVA) is valid, when assume that the envelope of a forward-travelling wave pulse varies slowly in time and space compared to a period or wavelength. We can estimate the smallest envelope size in our configuration, around 13 nm \cite{Shafraniuk2015}, which are much larger than wavelength of electron in graphene, 0.74 nm \cite{Meyer2008}. Therefore, slowly envelope varying approximation is a good approximation in our paper. Slowly envelope varying approximation is very widely used method in the optics wave \cite{Yariv1973, Longhi2005, Huang2014} and the slowly envelope varying approximation also can work for the electron wave function \cite{Burt1999}. We can ignore the first term. The third term and fourth terms can be cancelled out each other, based on Eq. (A2).






\end{document}